\documentstyle{article}
\input math_macros
\font\tenbf=cmbx10
\font\tenrm=cmr10
\font\tenit=cmti10
\font\elevenbf=cmbx10 scaled\magstep 1
\font\elevenrm=cmr10 scaled\magstep 1
\font\elevenit=cmti10 scaled\magstep 1

\renewenvironment{thebibliography}[1]
 { \elevenrm
   \begin{list}{\arabic{enumi}.}
    {\usecounter{enumi} \setlength{\parsep}{0pt}
     \setlength{\itemsep}{3pt} \settowidth{\labelwidth}{#1.}
     \sloppy
    }}{\end{list}}

\begin{document}
\begin{center}{{\tenbf GLOBAL QUANTIZATION OF VACUUM ANGLE
 AND MAGNETIC \\
\vglue 10pt
MONOPOLES AS A NEW SOLUTION TO THE STRONG CP PROBLEM
\footnote{JSUHEP921201, Work supported in Theoretical Physics Group,
Lawrence Berkeley Laboratory by a science consortium award}\\}
\vglue 5pt
\vglue 1.0cm
{\tenrm {\bf HUAZHONG ZHANG} \\}
\baselineskip=13pt
{\tenit Department of Physics and Atmospheric Sciences,
Jackson State University,\\}
\baselineskip=12pt
{\tenit P.O.Box 17660, Jackson, MS 39217, USA\\}

\vglue 0.8cm
{\tenrm ABSTRACT}}
\end{center}
\vglue 0.3cm
{\rightskip=3pc
 \leftskip=3pc
 \tenrm\baselineskip=12pt
 \noindent
The non-perturbative solution to the strong CP problem with
magnetic monopoles as originally proposed by the author is
described. It is shown that the gauge orbit space with gauge
potentials and gauge tranformations restricted on the space
boundary and the globally well-defined gauge subgroup in
gauge theories with a $\theta$ term has a monopole structure
if there is a magnetic monopole in the ordinary space. The
Dirac's quantization condition then ensures that the vacuum
angle $\theta$ in the gauge theories must be quantized to
have a well-defined physical wave functional. The
quantization rule for $\theta$ is derived as $\theta=0,
2\pi/n~(n\neq 0)$ with n being the topological charge of the
magnetic monopole. Therefore, the strong CP problem is
automatically solved with the existence of a magnetic
monopole of charge $\pm 1$ with $\theta=\pm 2\pi$. This is
also true when the total magnetic charge of monopoles are
very large ($|n|\geq 10^92\pi$).
The fact that the strong CP violation can be only so small
or vanishing may be a signal for the existence of magnetic
monopoles and the universe is open.
\vglue 0.6cm}
{\elevenbf\noindent 1. Introduction and Summary of the Main
Results}
\baselineskip=14pt
\vglue 0.5cm
\elevenrm
Yang-Mills theories$^1$ and their non-perturbative effects
have played one of the most important roles in particle
physics. It is known that, in non-abelian gauge theories a
Pontryagin or $\theta$ term,
\begin{equation}
{\cal L}_{\theta}=\frac{\theta}{32{\pi}^2}\epsilon^{\mu\nu\lambda\
sigma}F^
a_{\mu\nu}F^a_{\lambda\sigma},
\end{equation}
can be added to the Lagrangian density of the system due to
instanton$^2$
effects in gauge theories. This term can induce CP
violations for an abitrary value of $\theta$. Especially,
such an effective $\theta$ term in QCD may induce CP
violations in strong interactions. In our discussions
relevant to QCD, $\theta$ is simply used to denote
$\theta+arg(detM)$ effectively with M being the
quark mass matrix, when the effects of electroweak
interactions are included. However, the experimental results
on the neutron electric dipole moment strongly limit the
possible values of the $\theta$ in QCD ($\leq 10^{-9}$,
modulo 2$\pi$ for example). This is the well-known strong CP
problem. One of the most interesting understanding of the
strong CP problem has been
the assumption of an additional Peccei-Quinn $U(1)_{PQ}$
symmetry$^4$, but the observation has not given$^3$ evidence
for the axions$^5$ needed in this approach. Thus the other
possible solutions to this problem are of fundamental
interest.

Recently, a non-perturbative solution to the Strong CP
problem with magnetic monopoles has been proposed originally
by the author$^{6}$. In our solution$^{6}$, it is proposed
that the vacuum angle with magnetic monopoles must be
quantized. Our quantization rule is derived essentially by
two different methods. This is given by $\theta=0$, or
$\theta=2\pi N/n~ (n\neq 0)$ with integer n being the
relevant topological charge of the magnetic monopole and N
may be fixed as 1 in the method 1 and is an arbitrary
integer in the method 2. The first method$^1$ is to show the
existence of a monopole structure in the relevant gauge
orbit space in Scherodinger formulation$^{7,8}$, and using
the Dirac quantization rule for having a well-defined wave
funtional. The second method is to show that there exist
well-defined gauge transformations which will ensure the
quantization of $\theta$ by the constraints of Gauss's law
due to the non-abelican electric charges carried by the
magnetic monopoles proportional to $\theta$ as noted in Ref.
21 and generalized in Ref.22 to the non-abelian case for the
generalized magnetic monopoles$^{17}$.

Therefore, we conclude that strong CP problem can be solved
due to the quantization of $\theta$ in the presence of
magnetic monopoles, for example monopoles of topological
charge $n=\pm 1$ with
$\theta=\pm 2\pi$, or $n\geq 2\pi 10^9$ with $\theta\leq
10^{-9}$.
Moreover, the existence of non-vanishing magnetic flux
through the space boundary implies that the universe must be
open. In this note, we will briefly describe and review our
solution to the strong CP problem with magnetic monopoles
with the first method.

\vglue 0.6cm
{\elevenbf\noindent 2. Quantization Condition on $\theta$
and Solution to the Strong CP Problem}
\baselineskip=14pt
\vglue 0.5cm
\elevenrm

The main idea of our discussions is based on the follows. A
wave functional in the gauge orbit space corresponds to a
cross section$^{12}$ of the relevant fiber bundle for the
theory. Topologically, if there is a non-vanishing gauge
field as the curvature in the gauge orbit space, then the
flux of the curvature through a closed surface in the gauge
orbit space must be quantized to have
a cross section$^{7-8}$. Physically, this is equivalent to
say that the magnetic flux through the closed surface must
be quantized according to the Dirac quantization condition
in order to have a well-defined wave functional in the
quantum theory.

In this method, we will extend the method of Wu and Zee in
Ref.7 for the discussions of the effects of the Pontryagin
term in pure Yang-Mills theories in the gauge orbit spaces
in the Schrodinger formulation. This formalism has also been
used with different methods to derive the mass parameter
quantization in three-dimensional Yang-Mills theory with
Chern-Simons term$^{7-8}$. It is shown in Ref. 7 that the
Pontryagin term induces an abelian background field or an
abelian structure in the gauge configuration space of the
Yang-Mills theory. In our discussions, we will consider the
case with the existence of a magnetic monopole. We will show
that magnetic monopoles$^{9-10}$ in space will induce an
abelian gauge field with non-vanishing field strength in
gauge configuration space, and magnetic flux through a
two-dimensional sphere in the induced gauge orbit space on
the space boundary is non-vanishing. Then, Dirac
condition$^{9-10}$ in the
corresponding quantum theories leads to the result that the
relevant vacuum angle $\theta$ must be quantized as
$\theta=2\pi/n$ with n being the
topological charge of the monopole to be generally defined.
Therefore, the strong CP problem can be solved with the
existence of magnetic monopoles.

We will now consider the Yang-Mills theory with the
existence of a magnetic monopole at the origin. Our
derivation applies generally to a gauge theory with an
arbitrary simple gauge group or a U(1) group outside the
monopole. This gauge group under consideration may be
regarded as a factor group in the exact gauge group of a
grand unification theory. Note that there can be Higgs field
and unification gauge fields confined inside the monopole
core, which will be ignored in our discussion outside the
monopoles.

As we will see that an interesting feature in our
derivation is that we will use the Dirac quantization
condition both in the ordinary space and restricted gauge
orbit space to be defined. The Lagrangian
of the system is given by
\begin{equation}
{\cal L}={\int}d^4x\{-\frac{1}{4}F^a_{\mu\nu}F^{a\mu\nu}+\frac{\theta}{32\pi^2}
\epsilon^{\mu\nu\lambda\sigma}F^a_{\mu\nu}F^{a\lambda\sigma}\}.
\end{equation}
We will use the Schrodinger formulation and the Weyl gauge
$A^0=0$. The conjugate
momentum corresponding to $A^a_i$ is given by
\begin{equation}
\pi^a_i=\frac{\delta{\cal
L}}{\delta\dot{A}^a_i}=\dot{A}^a_i+\frac{\theta}
{8\pi^2}\epsilon_{ijk}F^a_{jk}.
\end{equation}
In the Schrodinger formulation, the system is similar to the
quantum system
of a particle with the coordinate $q_i$ moving in a gauge
field $A_i(q)$ with
the correspondence$^{6-7}$
\begin{eqnarray}
q_i(t)\rightarrow A^a_i({\bf x},t),\\
A_i(q)\rightarrow {\cal A}^a_i({\bf A}({\bf x})),
\end{eqnarray}
where
\begin{eqnarray}
{\cal A}^a_i({\bf A}({\bf x}))=\frac{\theta}{8\pi^2}\epsilon_{ijk}F^a_{jk}.
\end{eqnarray}
Thus there is a gauge structure with gauge potential ${\cal
A}$ in this formalism within a gauge theory with the
${\theta}$ term included. Note that in our discussion with
the presence of a magnetic monopole, the gauge
potential ${\bf A}$ outside the monopole generally need to
be understood as well defined in each local coordinate
region. In the overlapping regions, the
separate gauge potentials can only differ by a well-defined
gauge transformation$^{10}$. In fact, single-valuedness of
the gauge function
corresponds to the Dirac quantization condition$^{10}$. For
a given r, we can
choose two extended semi-spheres around the monopole, with
$\theta
\in[\pi/2-\delta,\pi/2+\delta] (0<\delta<\pi/2)$ in the
overlapping region,
where the $\theta$ denotes the $\theta$ angle in the
spherical polar coordinates. For convenience, we will use
differential forms$^{10}$ in our
discussions, where $A=A_idx^i, F=\frac{1}{2}F_{jk}dx^jdx^k,$
with $F=dA+A^2$
locally. For our purpose to discuss about the effects of the
abelian gauge
structure on the quantization of the vacuum angle, we will
now briefly clarify the relevant topological results needed,
then we will realize the topological results explicitly.

With magnetic monopoles, we need to generalize the gauge
orbit space of ordinary gauge theories to include the space
boundary which is noncontractible with non-vanishing
magnetic flux quantized according Dirac quantization
condition. With the constraint of Gauss' law, the quantum
theory in the finite space region in this formalism is
described in the usual gauge orbit space ${\cal U}$/${\cal
G}$. The ${\cal U}$ is the space of well-defined gauge
potentials and ${\cal G}$ denotes the space of continuous
gauge transformations with gauge functions mapping the space
boundary to a single point in the gauge group. Due to the
exitence of magnetic monopoles, the gauge transformations on
the space boundary $S^2$ can be non-trivial, the physical
effects of the well-defined gauge transformations need to be
considered. As it is known that$^{12}$, only the gauge
transformations generated by the generators commuting with
magnetic charges may be well-defined globally. On the space
boundary, ${\cal U}$ will also be used to denote the induced
gauge configuration space with gauge potentials restricted
on the space boundary, and ${\cal G}$ also denotes the
continuous gauge transformations restricted on the space
boundary and well-defined gauge subgroup. Then we will call
corresponding ${\cal U}$/${\cal G}$ as restricted gauge
orbit space. Collectively, they will be called as the usual
space for the finite coordinate space region and the
restricted space on the space boundary. There should not be
confusing for the notations used both for the usual spaces
and restricted spaces. As we will see that the magnetic
charges up to a conjugate transformation are in a Cartan
subalgebra of the gauge group, then on the space boundary
$S^2$, we need to consider a well-defined gauge subgroup
$G=U(1)$ for the quantization of $\theta$. Similar to the
usual gauge orbit space on the compactified coordinate space
by restricting to gauge functions mapping the space boundary
to a single point in the guage group, the restricted gauge
orbit space is well-defined since the space boundary $S^2$
is compact.

     Note that the physical meaning of the restricted gauge
orbit space can be understood as follows. Let
$\Psi_{phys}({\bf A({\bf x})})$ denote the physical wave
functional and $\Psi_{phys}({\bf A({\bf x})})\mid_{S^2}$ be
its restriction on the space boundary $S^2$ which actually
only depends on the direction of ${\bf x}$. Then, the
$\Psi_{phys}\mid_{S^2}$ must be invariant under the gauge
transformations well-defined on the entire space boundary.
Namely the $\Psi_{phys}\mid_{S^2}$ is defined in the
restricted gauge orbit space.
However, in the finite space region, the $\Psi_{phys}({\bf
A({\bf x})})$ for finite ${\bf x}$ is only required to be
invariant under the gauge transformations with gauge
function going to the identity at the spatial infinity.
Namely, it is defined the usual gauge orbit space. The
entire $\Psi_{phys}$ is then well-defined in the generalized
gauge orbit space as described.

Now consider the following exact homotopy sequence$^{13}$
both for the ususal and restricted spaces:
\begin{equation}
\Pi_N({\cal U})\stackrel{P_*}{\longrightarrow}\Pi_N({\cal
U}/{\cal G})
\stackrel{\Delta_*}{\longrightarrow}\Pi_{N-1}({\cal
G})\stackrel{i_*}
{\longrightarrow}\Pi_{N-1}({\cal U}) ~ (N\geq 1).
\end{equation}
Note that homotopy theory has also been used to study the
global
gauge anomalies $^{14-16}$, especially by using
extensively the exact homotopy sequences and in terms of
James numbers of
Stiefel manifolds$^{17}$. One can easily see that
${\cal U}$ is topologically trivial,
thus $\Pi_N({\cal U})=0$ for any N. Since
the interpolation between any two gauge potentials $A_1$ and
$A_2$
\begin{equation}
A_t=tA_1+(1-t)A_2
\end{equation}
for any real t is in ${\cal U}$ (Theorem 7 in Ref.10, and
Ref.7). since $A_t$ is
transformed as a gauge potential in each local coordinate
region, and in an
overlapping region, both $A_1$ and $A_2$ are gauge
potentials may be defined up
to a gauge transformation, then $A_t$ is a gauge potential
which may be defined
up to a gauge transformation, namely, $A_t\in{\cal U}$.
Thus, we have
\begin{equation}
0\stackrel{P_*}{\longrightarrow}\Pi_N({\cal U}/{\cal G})
\stackrel{\Delta_*}{\longrightarrow}\Pi_{N-1}({\cal
G})\stackrel{i_*}
{\longrightarrow}0 ~ (N\geq 1).
\end{equation}
This implies that
\begin{equation}
\Pi_N({\cal U}/{\cal G})\cong\Pi_{N-1}({\cal G})~ (N\geq 1).
\end{equation}

As we will show that in the presence of a magnetic monopole,
the topological
properties of the system are drastically different. This
will give important
consequences in the quantum theory.
In fact, the topological properties of the restricted gauge
orbit spaces are relevant for our purpose since as we will
see that only the integrals on the space
boundary $S^2$ are relevant in the quantization equation for
the $\theta$. Now for
the restricted spaces, the main topological result we will
use is given by
\begin{equation}
\Pi_2({\cal U}/{\cal G})\cong\Pi_1({\cal G}) =\Pi_1(G)\oplus\Pi_3(G),
\end{equation}
for a well-defined gauge subgroup G. As we will see that in
the relevant case of $G=U(1)$ for our purpose $\Pi_3(G)=0$.
The condition $\Pi_2({\cal U}/{\cal G})\neq 0$ corresponds
to the existence of a magnetic monopole in the restricted
gauge orbit space. We will first show that in this case
${\cal F}\neq 0$, and then demonstrate explicitly that the
magnetic flux $\int_{S^2}\hat{\cal F}\neq 0$ can be
nonvanishing in the restricted gauge orbit space, where
$\hat{\cal F}$ denotes the projection of $\cal F$ into the
restricted gauge orbit space.

Denote the differentiation with respect to
space variable ${\bf x}$ by d, and the differentiation with
respect to
parameters $\{t_i\mid i=1,2...\}$ which {\bf A}({\bf x}) may
depend on in the
gauge configuration space by $\delta$, and assume
$d\delta+\delta d$=0. Then,
similar to $A=A_{\mu}dx^{\mu}$ with $\mu$ replaced by a, i,
${\bf x}$,
$A=A^a_iL^adx^i, F=\frac{1}{2}F^a_{jk}L^adx^jdx^k$ and
$tr(L^aL^b)=-\frac{1}{2}{\delta}^{ab}$ for a basis
$\{L^a\mid a=1, 2,...,rank(G)\}$ of the Lie algebra of the
gauge group G,
the gauge potential in the gauge configuration space is
given by
\begin{equation}
{\cal A}=\int d^3x{\cal A}^a_i({\bf A}({\bf x}))\delta
A^a_i(\bf x).
\end{equation}
Using Eq.(6), this gives
\begin{equation}
{\cal A}=\frac{\theta}{8\pi^2}\int d^3x\epsilon_{ijk}F^a_{jk}({\bf x})\delta
A^a_i({\bf x})=-\frac{\theta}{2\pi^2}\int_M tr(\delta AF),
\end{equation}
with M being the space manifold. With $\delta F=-D_A(\delta
A)=
-\{d(\delta A)+A\delta A-\delta AA\}$, we have topologically
\begin{equation}
{\cal F}=\delta{\cal A}=\frac
{\theta}{2\pi^2}\int_Mtr[\delta AD_A(\delta A)]
=\frac {\theta}{4\pi^2}\int_Mdtr(\delta A\delta A)
=\frac {\theta}{4\pi^2}\int_{\partial M}tr(\delta A\delta
A).
\end{equation}
Usually, one may assume $A\rightarrow 0$ faster than 1/r as
$\bf x\rightarrow 0$
, then$^7$ this would give ${\cal F}=0$. However, this is
not the case in the
presence of a magnetic monopole. Asymptotically, a monopole
may generally give
a field strength of the form$^{9-10,17}$
\begin{equation}
F_{ij}=\frac{1}{4\pi r^2}\epsilon_{ijk}({\bf {\hat
r}})_kG({{\bf \hat r}}),
\end{equation}
with $\bf {\hat r}$ being the unit vector for {\bf r}, and
this gives
$A\rightarrow O(1/r)$ as $\bf x\rightarrow 0$. Thus, one can
see
easily that a magnetic monopole can give a nonvanishing
field strength $\cal F$
in the gauge configuration space. To evaluate the $\cal F$,
one needs to
specify the space boundary $\partial M$ in the presence of a
magnetic monopole.
we now consider the case that the magnetic monopole does not
generate a
singularity in the space. In fact, this is so when monopoles
appear as a smooth
solution of a spontaneously broken gauge theory similar to
't Hooft Polyakov
monopole$^9$. For example, it is known that$^{18}$ there are
monopole solutions
in the minimal SU(5) model. Then, the space boundary may be
regarded as a large
2-sphere $S^2$ at spatial infinity. For our purpose, we
actually only need to
evaluate the projection of $\cal F$ into the gauge orbit
space.

In the gauge orbit space, a gauge potential can be written
in the form of
\begin{equation}
A=g^{-1}ag+g^{-1}dg,
\end{equation}
for an element a $\in{\cal U}/{\cal G}$ and a gauge function
$g\in{\cal G}$.
Then the projection of a form into the gauge orbit space
contains only terms
proportional to $(\delta a)^n$ for integers n. We can now
write
\begin{equation}
\delta A=g^{-1}[\delta a-D_a(\delta gg^{-1})]g.
\end{equation}
Then we obtain
\begin{equation}
{\cal A}=-\frac{\theta}{2\pi^2}\int_M tr(f\delta a)
+\frac{\theta}{2\pi^2}\int_M tr[fD_a(\delta gg^{-1})],
\end{equation}
where $f=da+a^2$. With some calculations, this can be
simplified as
\begin{equation}
{\cal A}=\hat{\cal A}
+\frac{\theta}{2\pi^2}\int_{S^2}tr[f\delta gg^{-1}],
\end{equation}
where
\begin{equation}
\hat{\cal A}=-\frac{\theta}{2\pi^2}\int_M tr(f\delta a),
\end{equation}
is the projection of $\cal A$ into the gauge orbit space.
Similarly, we have
\begin{equation}
{\cal F}=\frac {\theta}{4\pi^2}\int_{S^2}
tr\{[\delta a-D_a(\delta gg^{-1})][\delta a-D_a(\delta gg^{-
1})]\}
\end{equation}
or
\begin{equation}
{\cal F}=\hat{\cal F}-\frac {\theta}{4\pi^2}\int_{S^2}
tr\{\delta aD_a(\delta gg^{-1})+D_a(\delta gg^{-1})\delta a
-D_a(\delta gg^{-1})D_a(\delta gg^{-1})\},
\end{equation}
where
\begin{equation}
\hat{\cal F}=\frac {\theta}{4\pi^2}\int_{S^2}tr(\delta
a\delta a).
\end{equation}
Now all our discussions will be based on the restricted
spaces.
To see that the flux of $\hat{\cal F}$ through a closed
surface in the
restricted gauge orbit space ${\cal U}/{\cal G}$ can be
nonzero, we will
construct a 2-sphere in it. Consider an element $a\in{\cal
U}/{\cal G}$, and a
loop in $\cal G$. The set of all the gauge potentials
obtained by all the gauge
transformations on $a$ with gauge functions on the loop then
forms a loop
$C^1$ in the gauge configurations space $\cal U$. Obviously,
the $a$ is the
projection of the loop $C^1$ into ${\cal U}/{\cal G}$.
Now since $\Pi_1({\cal U})=0$ is trivial, the loop $C^1$ can
be continuously
extented to a two-dimensional disc $D^2$ in the $\cal U$
with
$\partial D^2=C^1$, then obviously, the projection of the
$D^2$ into the
gauge orbit space is topologically a 2-sphere
$S^2\subset{\cal U}/{\cal G}$.
With the Stokes' theorem in the gauge configuration space,
We now have
\begin{equation}
\int_{D^2}{\cal F}=\int_{D^2}\delta{\cal A}=\int_{C^1}{\cal
A}.
\end{equation}
Using Eqs.(19) and (24) with $\delta a=0$ on $C^1$, this
gives
\begin{equation}
\int_{C^1}{\cal A}
=\frac{\theta}{2\pi^2}tr\int_{S^2}\int_{C^1}[f\delta gg^{-
1}].
\end{equation}
Thus, the projection of the Eq.(26) to the gauge orbit space
gives
\begin{equation}
\int_{S^2}\hat{\cal F}
=\frac{\theta}{2\pi^2}tr\int_{S^2}\{f\int_{C^1}\delta gg^{-
1}\},
\end{equation}
where note that in the two $S^2$ are in the gauge orbit
space and the ordinary
space respectively. We have also obtained this by verifying
that
\begin{equation}
\int_{D^2}tr\int_{S^2}
tr\{\delta aD_a(\delta gg^{-1})+D_a(\delta gg^{-1})\delta a
-D_a(\delta gg^{-1})D_a(\delta gg^{-1})\}=0,
\end{equation}
or the projection of $\int_{D^2}{\cal F}$ gives
$\int_{S^2}\hat{\cal F}$.

In quantum theory, Eq.(26) corresponds to the topological
result
$\Pi_2({\cal U}/{\cal G})\cong\Pi_{1}({\cal G})$ on the
restricted spaces.
The discussion about the Hamiltonian equation in the
schrodinger formulation
will be similar to that in Refs.7 and 8 including the
discussions for the
three-dimensional Yang-Mills theories with a Chern-Simons
term. We need the Dirac quantization condition to have a
well-defined wave functional in the formalism. In the gauge
orbit space,
the Dirac quantization condition gives
\begin{equation}
\int_{S^2}\hat{\cal F}=2\pi k,
\end{equation}
with k being integers. The Dirac quantization condition in
the gauge orbit space will be clarified shortly. Now let $f$
be the field strength 2-form for the
magnetic monopole. The quantization condition is now given
by$^{17}$
\begin{equation}
exp\{\int_{S^2}f\}=exp\{G_0\}=exp\{4\pi\sum_{i=1}^{r}\beta^{
i}H_{i}\}\in Z.
\end{equation}
Where $G_0$ is the magnetic charge up to a conjugate
transformation by a group
element, $H_i$ (i=1, 2,...,r=rank(G)) form a basis for the
Cartan subalgebra of
the gauge group with simple roots $\alpha_i$ (i=1,2,...,r).
We need non-zero topological value to obtain quantization
condition for $\theta$. As it is known from Ref.12, only the
gauge transformations commuting with the magnetic charges
can be globally well-defined, only those gauge
transformations can be used for determining the global
topological quantities. Consider g({\bf x},t) in the well-
defined U(1) gauge subgroup commutative with the magnetic
charges on the $C^1$
\begin{equation}
g({\bf x}, t)\mid_{x\in S^2}=exp\{4\pi
mt\sum_{i,j}\frac{(\alpha_i)^jH_j}{<\alpha_i,\alpha_i>}\},
\end{equation}
with m being integers and $t\in [0,1]$. In fact, m should be
identical to k according to our topological result
$\Pi_2({\cal U}/{\cal G})\cong\Pi_{1}({\cal G})$. The k and
m are the topological numbers on each side. Thus, we obtain
in the case of non-vanishing vacuum angle $\theta$
\begin{equation}
\theta=\frac{2\pi}{n}~(n\neq0).
\end{equation}
Where we define generally the topological charge of the
magnetic monopole as
\begin{equation}
n=-2<\delta,\beta'>
\end{equation}
which must be an integer$^{17}$. Where
\begin{equation}
\delta'=\sum_i\frac{2\alpha_i}{<\alpha_i,\alpha_i>},
\end{equation}
the minus sign is due to our normalization convention for
Lie algebra generators. Note that the parameter t of
g({\bf x},t) in
eq.(30) may be regarded as the time parameter topologically
when the time evolution is included, the two end points of
the closed loop then correspond to the time infinities. The
g({\bf x},t) is not a constant in the entire spacetime, and
does not generate a Nother symmetry, its invariance will not
eliminate any charged states.

Therefore, we conclude that in the presence of magnetic
monopoles with
topological charge $\pm 1$, the vacuum angle of non-abelian
gauge theories
must be $\pm 2\pi$, the existence of such magnetic monopoles
gives a solution
to the strong CP problem. But CP cannot be exactly conserved
in this case since
$\theta=\pm 2\pi$ correspond to two different monopole
sectors. The existence
of many monopoles can ensure $\theta\rightarrow 0$, and the
strong CP problem
may also be solved. In this possible solution to the strong
CP problem with
$\theta\leq 10^{-9}$, the total magnetic charges present are
$|n|\geq 2\pi 10^{9}$. This may possibly be within the
abundance allowed by
the ratio of monopoles to the entropy$^{19}$, but with the
possible existence
of both monopoles and anti-monopoles, the total number of
magnetic monopoles
may be larger than the total magnetic charges. Generally,
one needs to ensure
that the total number is consistent with the experimental
results on the
abundance of monopoles. The $n=\pm 2$ may also possibilely
solve CP if it is
consistent with the experimental observation.

Note that we only considered non-singular magnetic monopole
in the space. For 't Hooft Polykov monopole,
the full gauge group inside the monopole is simply
connected, it will not give any boundary contribution to the
term in Eq.(26). However, outside the monopole, the gauge
symmetry is spontaneously broken, it is known that the
unbroken gauge group cannot be simply connected to have
monopole solutions. For example, in SU(5) model, inside the
monopole, SU(5) is simply connected; outside the monopole
the exact gauge group G=SU(3)xU(1) satisfies $\Pi_1(G)=Z$.
We expect that in general, the GUT  monopoles are smooth
solutions, and therefore cannot have a mathematical boundary
at a given short distance around the monopole relevant to
our boundary contribution. Therefore, the realistic world
meet the condition to have our solution to the strong CP
problem.

The effect of a term proportional to
$\epsilon^{\mu\nu\lambda\sigma}F_{\mu\nu}
F_{\lambda\sigma}$ in the presence of magnetic charges was
first considered$^{20}$ relevant to chiral symmetry.
The effect of a similar U(1) $\theta$ term was discussed for
the purpose of
considering the induced electric charges$^{21}$ as quantum
excitations of dyons associated with the 't Hooft Polyakov
monopole and generalized magnetic
monopoles$^{16,21}$. Note that since our solution needs non-
vanishing magnetic flux through the space boundary, this
implies that only an open universe can be consistent with
our solution. Note that the relevance to the $U_A(1)$
problem is discussed in Ref. 23.

{\elevenbf \noindent  Acknowledgements \hfil}
\vglue 0.4cm
The author would like to express his gratitude to Y. S. Wu
and A. Zee for valuable discussions. He is also grateful to
O. Alvarez and the Theoretical Physics Group at the Lawrence
Berkeley Laboratory for their hospitality during the major
completion of the work.
\vglue 0.5cm
{\elevenbf\noindent References \hfil}
\vglue 0.4cm

\vglue 0.5cm
Email: ZHANGHZ@SSCVX1.SSC.GOV
\end{document}